\documentstyle[twoside,fleqn,espcrc2]{article}

\title{Standard Solar Models}

\author{John N. Bahcall\address{Institute for Advanced Study,
        Princeton, NJ 08540}}%

\begin{document}
\input psfig

\begin{abstract}
I review recent developments that affect standard solar model
predictions of solar neutrino fluxes.
\end{abstract}

\maketitle

\section{INTRODUCTION}
A lot of progress has been made in understanding the robustness of
solar model predictions since Neutrino 96~\cite{neutrino96}.  
In this talk, I will
first summarize the new ingredients and then give the current
best-estimates. Then I will discuss the uncertainties in the
predictions. 

Many of the results given here are adopted from the recent BP98
paper~\cite{BP98}. As we shall see in Section~\ref{best}, there is
excellent agreement between standard solar models calculated by
different groups with different codes.

If you want to obtain the numerical data that are discussed in this
talk, you can copy  them from my Web site: http://www.sns.ias.edu/$\sim$jnb .

\section{NEW INGREDIENTS}
\label{new}

In this section, I will first summarize the new and relevant results
on nuclear fusion reactions and on the screening of nuclear reactions
and then summarize the situation with respect to neutrino cross
sections.  Finally, I will mention a few miscellaneous improvements
that have been made since Neutrino 96.

\subsection{Nuclear reaction cross sections}

In January, 1997, the Institute for Nuclear Theory (INT) hosted a workshop
devoted to determining the best estimates and
the uncertainties in the
most important solar fusion reactions. Thirty-nine experts in 
low energy nuclear
experiments and theory, representing many different
research groups and points of view,  participated in the workshop and
evaluated the existing experimental data and theoretical 
calculations. Their conclusions  have been
summarized in a detailed article authored jointly by the participants
and published by 
the Reviews of Modern Physics \cite{adelberger98}.  In general 
outline, the conclusions
of the INT workshop paper confirmed and strengthened 
previous standard analyses of
nuclear fusion rates, although in a few important  cases 
(for the ${\rm ^3He}(\alpha, \gamma){\rm ^7Be}$, 
${\rm ^7Be}(p, \gamma){\rm ^8B}$, and 
${\rm ^{14}N}(p, \gamma){\rm ^{15}O}$ reactions)
the estimated
uncertainties were determined to be larger than previously believed.

The largest change from what was used in the results presented at
Neutrino 96 is 
the lower ${\rm ^7Be}(p,\gamma){\rm ^8B}$  cross section adopted
by Adelberger et al.~\cite{adelberger98}.
Previously, most authors constructing standard solar models used
the Caltech (CIT) value for the $^8$B production cross
section \cite{CIT}.
The difference between the INT and the CIT estimates of the $^8$B
production cross section is due almost entirely to the decision by the
INT group to base their estimate on only one (the best documented) of
the six experiments analyzed by the CIT collaboration.

As we go alone, I will indicate how the principal predictions of solar
models depend upon the assumed $^8$B production cross section.

\subsection{Screening of nuclear reactions}

In one respect, the calculation of neutrino fluxes has simplified from
Neutrino 96 to Neutrino 98.  The rather complicated expressions  in the
literature for the
screening of nuclear fusion reactions by electrons  and ions have been
replaced by a simple analytic expression that was originally derived
by Salpeter~\cite{salpeter} for the case of ``weak screening'' only.

Gruzinov and Bahcall~\cite{gruzinov} employed 
a mean field formalism to calculate
the electron density  of the screening cloud 
using the appropriate density matrix equation
of quantum statistical mechanics.
Because of well understood physical effects that are 
included for the first time in this treatment,
the calculated enhancement of reaction rates
does not agree with the frequently used interpolation 
formulae. For the sun, screening effects cause only small
uncertainties in the predicted neutrino fluxes if the appropriate
Salpeter formula is used.

\subsection{Neutrino cross sections and energy spectra}

Improved   cross sections for neutrino absorption by 
gallium and by chlorine are now available as well as somewhat more
precise standard (undistorted) neutrino energy spectra.
I summarize the results here. 

A number of  authors do not use the best
available data for the neutrino cross sections and energy spectra and some
authors even give event rates (or SNU values) but 
do not say which cross sections and energy spectra 
 they use so that one cannot
interpret their results precisely.

I have calculated~\cite{bahcall97} neutrino absorption cross sections for ${\rm^{71}Ga}$
for all solar neutrino sources
with standard energy spectra, and for  laboratory sources of ${\rm^{51}Cr}$
and ${\rm^{37}Ar}$;  
the calculations  including, where appropriate,  
the thermal energy of  fusing
solar ions  and 
use improved nuclear and atomic data.
The ratio, $R$, of measured (in GALLEX and SAGE) to calculated ${\rm ^{51}Cr}$ capture rate
is $R = 0.95 \pm 0.07~{\rm (exp)}~+~^{+0.04}_{-0.03}~{\rm (theory)}$
and was discussed extensively at Neutrino 98 by Gavrin and by Kirsten.

I also calculated cross sections    for  specific
neutrino energies chosen so that a spline fit 
determines accurately 
the event rates in a gallium detector even if new physics
changes the energy spectrum of solar neutrinos. 
In order to make possible more precise analyses  of event rates for neutrino
scenarios which change the shape of the neutrino energy spectra from
individual neutrino sources, I  evaluated and presented, for the first time, theoretical
uncertainties for absorption cross sections at specific energies, as
well as for the standard (undistorted) neutrino energy spectra.
Also for use by  people doing neutrino oscillation
calculations, I calculated 
standard 
energy spectra for pp and CNO neutrino
sources and presented the results in Appendices and on my Web site.

I note in passing that
neutrino fluxes predicted by standard solar models, 
corrected for
diffusion,  have been in the 
range $120$ SNU to $141$ SNU since 1968~\cite{bahcall97}.

A group of us have recently redetermined the standard shape for the
$^8$B neutrino energy spectrum~\cite{boronspectrum}. The available
data all seem to be consistent with each other within rather small
uncertainties, so we were able to determine not only a best-fit energy
spectrum but also two extreme spectra that are different by what we
estimate is effectively $\pm 3\sigma$.  
The uncertainties include estimates of the radiative and forbidden
corrections.
This improved spectrum yields a
slightly different $^8$B absorption cross
section~\cite{boronspectrum}. 

I am somewhat nervous about the standard (undistorted) $^8$Be neutrino
energy spectrum  since it does depend upon
rather old data~\cite{boronspectrum}.  
It would be very good if the $\alpha$-particle
energy spectrum from $^8$Be decay could be
remeasured accurately in a new laboratory experiment.

The situation for neutrino-electron scattering is good; cross sections
are available~\cite{sirlin} that include electroweak radiative
corrections. 

\subsection{Miscellaneous improvements}

The standard solar model, BP98, that is discussed  in this report
includes somewhat improved radiative opacities calculated by the
Livermore National Laboratory group, the so-called OPAL96 
opacities~\cite{opacity}, 
and the improved OPAL equation of state \cite{eos}.

One improvement that I am rather proud of is new publicly
available software that I have made available on my Web site (under
Neutrino Software and Data) is a program to calculate solar neutrino
rates and uncertainties. The problem of calculating the uncertainties
in the predicted fluxes is somewhat complicated, especially since the
uncertainties are asymmetric for some of the important input
parameters. I decided to make this software publicly available (and
therefore polished it significantly) so that people could see
explicitly what uncertainties were included for each parameter and how
the uncertainties were combined. 

I have also polished my nuclear energy generation code. The
changes made in this code, although they probably took me altogether a
few weeks of programming and debugging 
time, ended up not changing neutrino flux predictions by more
than a percent.

\section{BEST-ESTIMATE FLUXES\\ AND EVENT RATES}
\label{best}

Table~1 gives the  neutrino fluxes and their 
uncertainties for our best standard solar model,  hereafter BP98.
As discussed in the previous section, the solar model
makes use of the INT nuclear reaction rates, 
recent (1996) Livermore
OPAL radiative opacities, 
the OPAL equation of state,   and 
electron and ion screening as determined by the 
recent density matrix calculation.

\begin{table}[t]
\baselineskip=16pt
\centering
\begin{minipage}{7.5cm}
\caption[]{Standard Model Predictions (BP98): 
solar neutrino fluxes and neutrino capture rates, with $1\sigma$
uncertainties from all sources (combined quadratically).\protect\label{bestestimate}}
\begin{tabular}{l@{\extracolsep{-4pt}}lcc}
\noalign{\smallskip}
\hline
\noalign{\smallskip}
Source&\multicolumn{1}{c}{Flux}&Cl&Ga\\
&\multicolumn{1}{c}{$\left(10^{10}\ {\rm cm^{-2}s^{-1}}\right)$}&(SNU)&(SNU)\\
\noalign{\smallskip}
\hline
\noalign{\smallskip}
pp&$5.94 \left(1.00^{+0.01}_{-0.01}\right)$&0.0&69.6\\
pep&$1.39 \times 10^{-2}\left(1.00^{+0.01}_{-0.01}\right)$&0.2&2.8\\
hep&$2.10 \times 10^{-7}$&0.0&0.0\\
${\rm ^7Be}$&$4.80 \times 10^{-1}\left(1.00^{+0.09}_{-0.09}\right)$&1.15&34.4\\
${\rm ^8B}$&$5.15 \times 10^{-4}\left(1.00^{+0.19}_{-0.14}\right)$&5.9&12.4\\
${\rm ^{13}N}$&$6.05 \times
10^{-2}\left(1.00^{+0.19}_{-0.13}\right)$&0.1&3.7\\
${\rm ^{15}O}$&$5.32 \times
10^{-2}\left(1.00^{+0.22}_{-0.15}\right)$&0.4&6.0\\
${\rm ^{17}F}$&$6.33 \times
10^{-4}\left(1.00^{+0.12}_{-0.11}\right)$&0.0&0.1\\
\noalign{\medskip}
&&\hrulefill&\hrulefill\\
Total&&$7.7^{+1.2}_{-1.0}$&$129^{+8}_{-6}$\\
\noalign{\smallskip}
\hline
\noalign{\smallskip}
\end{tabular}
\end{minipage}
\end{table}

\begin{figure*}[htb] 
\centerline{\psfig{figure=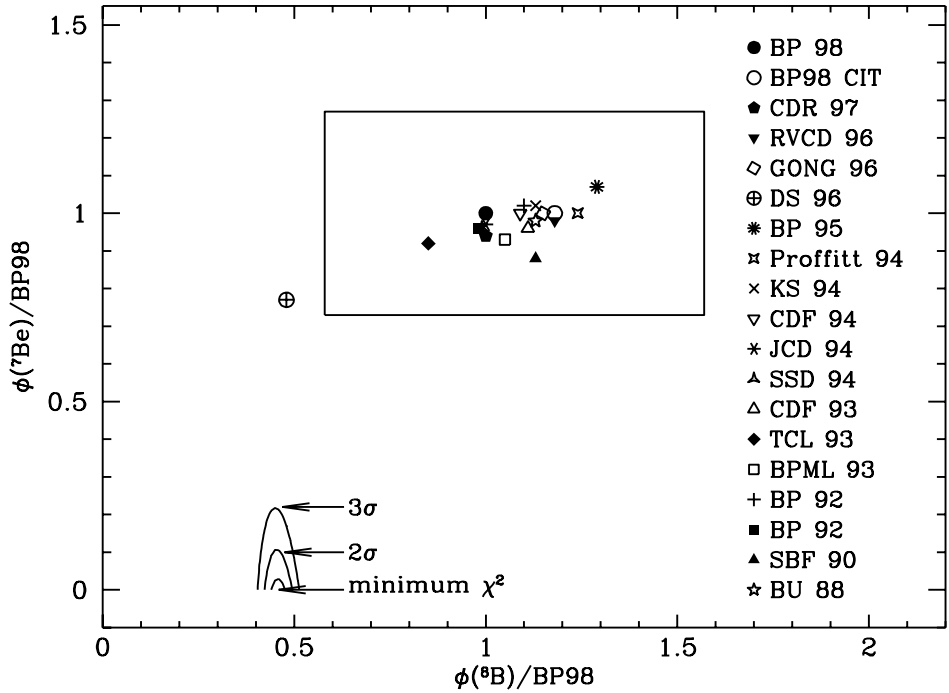,width=5in}}
\vglue-.2in
\caption[]{Predictions of standard solar models since 1988.
The figure
shows the predictions of $19$ standard solar models in the
plane defined by the ${\rm ^7Be}$ and ${\rm ^8B}$ neutrino fluxes. 
The abbreviations that are used in the figure to identify different
solar models are defined in the bibliographical item, Ref.~\cite{models}.
We include all standard solar models with which we are familiar that
were published in refereed journals
in the decade 1988-1998.
All of the
fluxes are normalized to the predictions of the 
Bahcall-Pinsonneault 98 solar model, BP98~\cite{BP98}.
The rectangular error box defines the $3\sigma$ error range of the
BP98 fluxes. The best-fit ${\rm ^7Be}$ neutrino flux is negative. At the
$99$\% C.L., there is no solution with all positive neutrino fluxes
if the fluxes of CNO neutrinos are arbitrarily set equal to
zero. There is no solution at the $99.9$\% C.L. if the CNO neutrinos
are fixed at their standard solar model values.
All of the standard model solutions lie  far from the best-fit
solution, even far from the $3\sigma$ contour.}
\label{fig:independent}
\end{figure*}

Figure~\ref{fig:independent}
displays the 
calculated  ${\rm ^7Be}$ and ${\rm
^8B}$ neutrino fluxes 
for all $19$ standard solar models with which we are
familiar which have been published in the last $10$ years in refereed
science journals.  
The fluxes are normalized by 
dividing each
published value by the flux from the BP98
solar model~\cite{BP98}; the abscissa is the
normalized ${\rm ^8B}$ flux and the ordinate
 is the normalized ${\rm ^7Be}$
neutrino flux.  The rectangular box shows the estimated 
$3\sigma$ uncertainties in the
predictions of the BP98 solar model.
The abbreviations, which  indicate references to individual models, are
identified in the caption of Figure~\ref{fig:independent}. 

All of the solar model results from different groups 
fall within the estimated 3$\sigma$
uncertainties in the BP98 analysis (with the exception of the
Dar-Shaviv model whose results have not been reproduced by other groups).
This agreement 
demonstrates the robustness of the predictions since the
calculations use different computer codes (which achieve varying
degrees of precision) and 
involve a variety of choices for the nuclear
parameters, the equation of state, the stellar radiative opacity, 
the initial heavy element abundances, and the physical processes
that are included.

The largest contributions to  the dispersion in values in
Figure~\ref{fig:independent} are due to the
choice of the normalization for  
$S_{17}$ (the production
cross-section factor for ${\rm ^8B}$ neutrinos) and the
inclusion, or non-inclusion, of element diffusion in the 
stellar evolution codes.
The effect in the plane of Fig.~\ref{fig:independent} 
of the normalization of
$S_{17}$ is shown by the difference between the point for BP98
(1.0,1.0), which was computed using the INT normalization, and the
point at (1.18,1.0) which corresponds to the BP98 result with the CIT
normalization.   

Helioseismological observations have shown ~\cite{neutrino96,prl97}
that diffusion is occurring
and must be included in solar models, so that the most recent models
shown in Fig.~\ref{fig:independent} now all include helium and heavy
element diffusion.
By comparing a large number of earlier models,
it was shown that 
all published standard  solar
models give the same results for solar neutrino fluxes to an accuracy
of better than 10\% if the same input parameters and physical
processes are included~\cite{BP92,BP95}.

The theoretical predictions in Table~1 
disagree with the observed neutrino event rates,
which are, see Ref.~\cite{experiments} and the results presented at
this conference by Lande, Gavrin, Kirsten, and Suzuki: 
$2.56 \pm 0.23$ SNU (chlorine), $72.2 \pm 5.6$ SNU (GALLEX and SAGE
gallium experiments), and $(2.44 \pm 0.10) \times 10^6 
{\rm cm^{-2} s^{-1} }$ ($^8$B flux from SuperKamiokande).

Bahcall, Krastev, and Smirnov~\cite{bks98} have compared the observed rates 
with the calculated, standard model values, combining  quadratically
the theoretical solar model and experimental
uncertainties, as well as the uncertainties in the neutrino cross
sections.
Since the GALLEX and SAGE experiments measure the same quantity, we
treat the weighted average rate in gallium as one experimental number.
We adopt the SuperKamiokande measurement as the most precise direct
determination of the higher-energy ${\rm ^8B}$ neutrino flux.

Using the predicted fluxes from the BP98
model, the $\chi^2$ for the fit to the three  experimental rates
(chlorine, gallium, and SuperKamiokande) is

\begin{equation}
\chi^2_{\rm SSM} \hbox{(3 experimental rates)} = 61\ .
\label{chitofit}
\end{equation}
The result given in Eq.~(\ref{chitofit}), which is approximately 
equivalent to  a
$20\sigma$ discrepancy,  is a quantitative expression 
of the fact that the standard model predictions do not fit the
observed solar neutrino measurements.

The principal differences between the results shown in Table~1 and the
results presented in our last systematic publication of calculated
solar neutrino fluxes \cite{BP95} is a  $1.3\sigma$ decrease in the 
$^8$B neutrino flux and $1.1 \sigma$ decreases
in the $^{37}$Cl and $^{71}$Ga capture rates.  These
decreases are due mainly  
to the lower ${\rm ^7Be}(p,\gamma){\rm ^8B}$  cross section adopted
by Adelberger et al. \cite{adelberger98}.
If we use, as in our recent previous publications,  
the Caltech (CIT) value for the $^8$B production cross
section \cite{CIT}, then the $^8$B flux is 
$\phi\left({\rm ^8B, CIT}\right) = 6.1^{+1.1}_{-0.9} 
\times 10^6~{\rm cm^{-2}s^{-1}}$,
$
\Sigma\left(\phi\sigma\right)_i \big\vert\lower10pt
\hbox{\scriptsize Cl,~CIT} =
8.8^{+1.4}_{-1.1}\ \ {\rm SNU} $,  and 
$\Sigma\left(\phi\sigma\right)_i 
\big\vert\lower10pt\hbox{\scriptsize\rm Gallium} = 131^{+9}_{-7}\ \
{\rm SNU}$, all 
of which are within ten percent of the
Bahcall-Pinsonneault 1995 best-estimates.

\begin{table*}[htb]
\baselineskip=16pt
\centering
\caption[]{Average uncertainties in neutrino fluxes and event rates 
due to different input data.  The flux uncertainties are expressed in
fractions of the total flux and the event rate uncertainties are
expressed in SNU.  The ${\rm ^7Be}$ electron capture rate causes an
uncertainty of $\pm 2\%$ \cite{be7paper} that affects only the ${\rm
^7Be}$ neutrino flux.  The average fractional uncertainties for
individual parameters are shown.}
\begin{tabular}{l@{\extracolsep{\fill}}ccccccccc}
\noalign{\smallskip}
\hline
\noalign{\smallskip}
$<$Fractional&pp&${\rm ^3He ^3He}$&${\rm ^3He ^4He}$&${\rm ^7Be} +
p$&$Z/X$&opac&lum&age&diffuse\\
uncertainty$>$&0.017&0.060&0.094&0.106&0.033&&0.004&0.004\\
\noalign{\smallskip}
\hline
\noalign{\smallskip}
Flux\\ \cline{1-1}
\noalign{\smallskip}
pp&0.002&0.002&0.005&0.000&0.002&0.003&0.003&0.0&0.003\\
${\rm ^7Be}$&0.0155&0.023&0.080&0.000&0.019&0.028&0.014&0.003&0.018\\
${\rm ^8B}$&0.040&0.021&0.075&0.105&0.042&0.052&0.028&0.006&0.040\\
\noalign{\medskip}
SNUs\\ \cline{1-1}
\noalign{\smallskip}
Cl&0.3&0.2&0.5&0.6&0.3&0.4&0.2&0.04&0.3\\
Ga&1.3&0.9&3.3&1.3&1.6&1.8&1.3&0.20&1.5\\
\noalign{\smallskip}
\hline
\noalign{\smallskip}
\end{tabular}
\end{table*}

\section{UNCERTAINTIES}
\label{uncertainties}

In this section, I will first discuss the formal uncertainties in the
solar model flux calculations and then review the strong constraints
that helioseismology places 
on perturbations of the standard solar model.

\subsection{Uncertainties in the flux calculations}
\label{fluxuncertainties}

Table~2 summarizes the uncertainties in the most important
solar neutrino fluxes and in the Cl and Ga event rates 
due to different nuclear fusion reactions (the
first four entries), the heavy element to hydrogen mass ratio (Z/X),
the radiative opacity, the solar luminosity, the assumed solar age,
and the helium and heavy element diffusion coefficients. 
The ${\rm ^{14}N} + p$ reaction causes a
0.2\% uncertainty in the predicted pp flux and a 0.1 SNU uncertainty
in the Cl (Ga) event rates.

The predicted event rates for
the chlorine and gallium experiments use recent improved calculations
of neutrino
absorption cross sections \cite{bahcall97,boronspectrum}. 
The uncertainty in the
prediction for the gallium rate is dominated by  uncertainties in the
neutrino absorption cross sections, $+6.7$ SNU ($7$\% of the predicted
rate) and $-3.8$ SNU ($3$\% of the predicted rate). The
uncertainties in the chlorine absorption cross sections cause an
error, $\pm 0.2$ SNU ($3$\% of the predicted rate), that is relatively
small compared to other
uncertainties in predicting the rate 
for this experiment.  For non-standard neutrino energy
spectra that result from new neutrino physics, the uncertainties in
the predictions 
for currently favored solutions (which reduce the contributions from
the least well-determined $^8$B neutrinos) will
in general be less than the values quoted here for standard spectra
and must be calculated using the appropriate cross section uncertainty
for each neutrino energy \cite{bahcall97,boronspectrum}.

\begin{figure*}[t]
\centerline{\psfig{figure=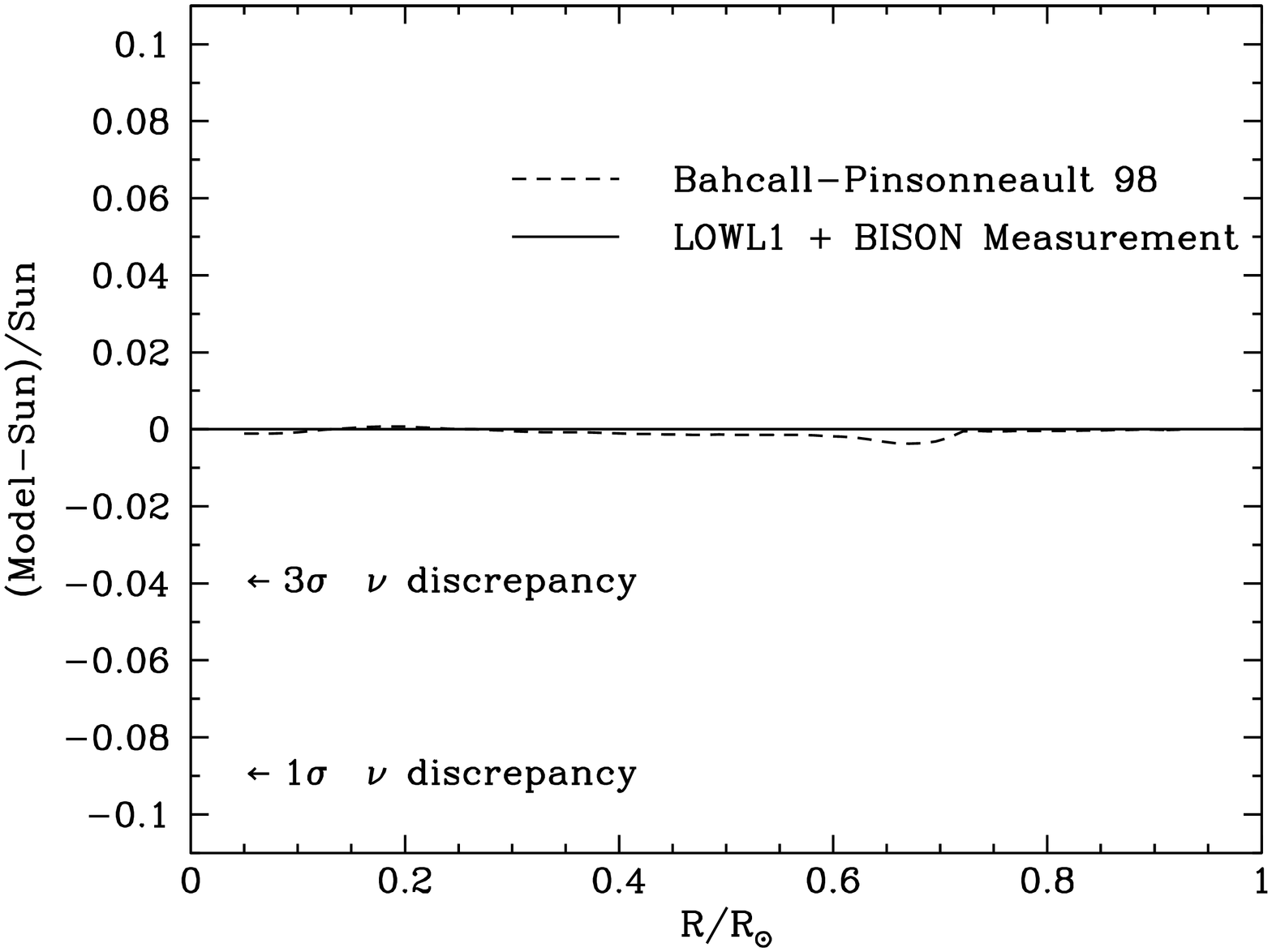,width=5in,angle=270}}
\vglue-.4in
\caption[]{Predicted versus Measured Sound Speeds. 
This figure shows
the excellent agreement between the calculated (solar model BP98, Model) 
 and the measured (Sun) sound
speeds, a fractional difference of $0.001$ rms for all 
speeds measured between $0.05 R_\odot$
and $0.95 R_\odot$. The vertical scale is chosen so as 
to emphasize that the fractional error is much smaller than generic
changes in the model, $0.03$ to $0.08$, that might
significantly affect the solar neutrino predictions.
\protect\label{modelsunnu}}
\end{figure*}

The nuclear fusion
uncertainties in Table~2 were taken  from Adelberger et
al. \cite{adelberger98}, 
the neutrino cross section uncertainties from \cite{bahcall97,boronspectrum}, 
the heavy element uncertainty was taken from
helioseismological measurements \cite{basu97}, the luminosity
and age uncertainties were adopted from BP95 \cite{BP95},  the
$1\sigma$ fractional uncertainty in the diffusion rate was taken to be
$15$\% \cite{thoul}, which is supported by helioseismological evidence
 \cite{prl97}, and the opacity uncertainty was determined by comparing
the results of fluxes computed using the older Los Alamos opacities
with fluxes computed using the modern Livermore opacities \cite{BP92}. 
 To include the effects of asymmetric errors, the now public-available
code for calculating rates and uncertainties (see discussion in
previous section)
was run with different input uncertainties and the results averaged.
The software contains a description of how each of the uncertainties
listed in Table~2 were determined and used.

The low energy cross section
of the ${\rm ^7Be} + p$ reaction is the most important quantity that must be
determined more accurately in order to decrease the error in the
predicted event rates in solar neutrino experiments.
The $^8$B neutrino flux that is measured by the
Kamiokande \cite{experiments}, 
Super-Kamiokande \cite{totsuka}, and SNO \cite{sno} 
experiments is, in all standard solar model calculations, 
directly proportional to the ${\rm ^7Be} + p$ cross section. 
If the $1\sigma$ uncertainty in this cross section can be reduced
by a factor of two to 5\%, then it will no longer be the limiting
uncertainty in predicting the crucial $^8$B neutrino flux (cf.~Table~2).

\subsection{How large an uncertainty does helioseismology suggest?}
\label{helioseismology}

Could the solar model calculations be wrong by enough to explain the
discrepancies between predictions and measurements for solar neutrino
experiments?  Helioseismology, which confirms predictions of the
standard solar model to high precision, suggests that the answer is
probably ``No.''

Figure~\ref{modelsunnu} shows the fractional differences 
between the most accurate 
available sound speeds
measured by helioseismology \cite{heliobest} and 
sound speeds calculated with our best
solar model (with no free parameters). 
The horizontal line corresponds to the hypothetical case in which the
model predictions exactly match the observed values.
The rms fractional 
difference between the
calculated and the measured sound speeds is $1.1\times10^{-3}$ for the
entire region over which the sound speeds are measured, $0.05R_\odot <
R < 0.95 R_\odot$.  In the solar core, $0.05R_\odot <
R < 0.25 R_\odot$ (in  which about $95$\% of the solar energy 
and neutrino flux is
produced in a standard model), the rms fractional 
difference between measured and
calculated sound speeds is $0.7\times10^{-3}$.

Helioseismological measurements also determine 
two other parameters that help characterize
 the outer part of the sun (far from
the inner region in which neutrinos are produced): the depth of the
solar convective zone (CZ), the region in the outer part of the sun
that is fully convective, and the present-day surface abundance by mass of
helium ($Y_{\rm surf}$).  The measured values, $R_{\rm CZ} =
(0.713 \pm 0.001) R_\odot$ \cite{basu95}, and $Y_{\rm surf} = 0.249 
\pm 0.003$ \cite{basu97},
are in satisfactory agreement with the values predicted by the solar
model BP98, namely, $R_{\rm CZ} =
0.714 R_\odot$, and $Y_{\rm surf} = 0.243$.
However, we shall see below that precision measurements of the sound
speed near the transition between the radiative interior (in which energy
is transported by radiation)  and the
outer convective zone (in which energy is transported by convection) 
reveal small discrepancies between the model
predictions and the observations in this region.

If solar physics were responsible for the solar neutrino problems, how
large would one expect the discrepancies to be between solar 
model predictions and
helioseismological observations?
The characteristic size of the discrepancies 
 can be estimated  using the results of the neutrino experiments and  
scaling laws for
neutrino fluxes and sound speeds.

All recently published solar models predict essentially 
the same fluxes from the fundamental
pp and pep reactions (amounting to $72.4$ SNU in gallium
experiments, cf.~Table~1), which are closely related to the solar
luminosity.  
Comparing the measured gallium rates (reported at Neutrino 98)  
and the standard predicted rate for the gallium
experiments, the $^7$Be flux must be reduced  by a factor $N$
if the disagreement is not to exceed $n$ standard deviations, where 
$N$ and $n$ satisfy  $72.4 + (34.4)/N = 72.2 + n \sigma$. For a $1
\sigma $ ($3\sigma$) disagreement, $ N = 6.1 (2.05)$.  
Sound
speeds scale like the square root of the local temperature divided by
the mean molecular weight
and the $^7$Be neutrino flux scales approximately as the $10$th power
of the temperature \cite{ulmer}. 
Assuming that the temperature changes are dominant,
agreement to within $1\sigma$ would
require fractional changes of order $0.09$ in sound speeds ($3\sigma$
could be reached with $0.04$ changes), if all model changes were in the
temperature\footnote{I have used in this calculation the GALLEX and
SAGE measured rates reported by Kirsten and Gavrin at Neutrino 98. The
experimental rates used in BP98 were not as precise and therefore
resulted in slightly less stringent constraints than those imposed
here. In BP98, we found that agreement to within $1\sigma$ with the
then available experimental numbers would
require fractional changes of order $0.08$ in sound speeds ($3\sigma$
could be reached with $0.03$ changes.)}.
This argument is conservative because it ignores the contributions
from the $^8$B and CNO neutrinos which contribute to the observed
counting rate (cf.~Table~1) and which, if included, would require an
even larger reduction of the $^7$Be flux. 

I  have chosen the vertical scale in Fig.~1 to be appropriate for
fractional differences between measured and predicted 
sound speeds that are of order $0.04$ to $0.09$ and 
that might therefore affect solar neutrino calculations. 
Fig.~1 shows that the characteristic agreement between solar model
predictions and helioseismological measurements is more than a factor
of $30$ better than would be expected if there were a solar model
explanation of the solar neutrino problems.

\section{DISCUSSION AND CONCLUSION}
\label{discussion}

Three decades of refining the input data
and the solar model calculations has led to a predicted standard model
event rate for
the chlorine experiment, $7.7$ SNU, which is very close to the
best-estimate value 
obtained in 1968 \cite{bahcall68}, which was $7.5$ SNU. The
situation regarding solar neutrinos is, however, completely different
now, thirty years later.
Four  experiments have confirmed the detection of solar neutrinos.
Helioseismological measurements show (cf.~Fig.~1) that hypothetical
deviations from the standard solar model that seem to be  required
by simple scaling laws  
to fit just the 
gallium solar neutrino results are at least a factor of $40$ larger
than the rms disagreement between the standard solar model predictions and the
helioseismological 
observations. This conclusion does not make use of 
the strong evidence which
points in the same direction from the chlorine, Kamiokande, and
SuperKamiokande experiments.

The improvement in  helioseismological measurements over
the past two years, from Neutrino 96 to Neutrino 98 (cf. Figure~2 of
the Neutrino~96 talk~\cite{neutrino96} with Figure~\ref{modelsunnu} of this
talk), has resulted in a five-fold improvement in the agreement with
the calculated standard solar model sound speeds and the measured
solar velocities! I believe that this improved agreement is yet
another reason to believe that standard solar models reliably predict
solar neutrino fluxes.

I am grateful to Y. Suzuki for special efforts that made possible my
attendance at Neutrino 98 and which enriched the scientific experience
of this extraordinarily exciting conference.
I acknowledge support from NSF grant \#PHY95-13835.

\end{document}